# A promise checked is a promise kept: Inspection Testing

Joachim Breitner
joachim@cis.upenn.edu
University of Pennsylvania
Philadelphia, PA, USA

**Abstract**

Occasionally, developers need to ensure that the compiler treats their code in a specific way that is only visible by *inspecting* intermediate or final compilation artifacts. This is particularly common with carefully crafted compositional libraries, where certain usage patterns are expected to trigger an intricate sequence of compiler optimizations – stream fusion is a well-known example.

The developer of such a library has to manually inspect build artifacts and check for the expected properties. Because this is too tedious to do often, it will likely go unnoticed if the property is broken by a change to the library code, its dependencies or the compiler. The lack of automation has led to released versions of such libraries breaking their documented promises.

This indicates that there is an unrecognized need for a new testing paradigm, *inspection testing*, where the programmer declaratively describes non-functional properties of an compilation artifact and the compiler checks these properties. We define inspection testing abstractly, implement it in the context of the Haskell Compiler GHC and show that it increases the quality of such libraries.

## 1 Introduction

The documentation of Haskell's popular text library by O'Sullivan and Harper [2017] makes a bold promise to its users:

> Most of the functions in this module are subject to fusion, meaning that a pipeline of such functions will usually allocate at most one Text value.

This is followed by an example of such a pipeline:

```
countChars :: ByteString –> Int
countChars = length . toUpper . decodeUtf8
```

The countChars function first decodes binary data into a value of type Text. Next, it changes every character to upper case, which can change the number of characters, producing a new value of type Text. Finally, it calculates the length of that value. The documentation promises that

> the two intermediate Text values will be optimized away, and the function will be compiled down to a single loop over the source ByteString.

This sounds like quite a feat. How can the author of the library – or a skeptical user – check if the feat is achieved?

Note that the promise is non-functional in nature: Even if the compiler fails to optimize the intermediate values away, the behavior of the code would not change. And, although runtime performance improvements are the motivation for fusion, simple benchmarking can not tell us with confidence whether fusion happened.

Instead, the author has to carefully *inspect* the output of the compiler and check that the promise holds. In this case, they would instruct GHC to dump a textual representation of its intermediate language Core and search for expressions that allocate a value of type Text. Presumably, that is what the library author did when they added the example above to the documentation.

Ideally, they would repeat this process after every change to the library, and with every new compiler release, to ensure that the promise still holds. But that is not practical! The result, however, is that at the time of writing, the promise is no longer fulfilled: One of the intermediate Text values does not get fused away.[1]

This anecdote shows that **non-functional properties of code need to be tested** if regressions are to be avoided. For such *inspection testing* to happen, four steps need to be taken:

1. Library authors need to notice when they make promises based only on manual inspection of compilation artifacts.

2. They need to be able to precisely describe the desired properties of the inspected artifact.

3. The compiler has to allow these properties to be declared, to check them during compilation.

4. Finally, the programmer needs to use this ability in their regression test suite.

In our example, the text maintainers took the first step by documenting their promises. As the second step, they would nail the property down more precisely and might say that "in the code of countChars, the type Text does not occur." We can implement such an analysis in the compiler (or in a compiler plugin) and come up with syntax to allow them to specify unit tests that look like

**inspect** countChar **contains no** Text

---

[1] May 2018, using text-1.2.3.0 and GHC-8.4, reported at https://github.com/haskell/text/issues/202.

Finally, they can include this declaration in their test suite. If a future change to the code breaks fusion, then this regression will be immediately spotted. Now the authors can make their promise with greater confidence, and without having to manually check it over and over again.

Elimination of intermediate data structures is just one example of a property that calls for inspection testing, but many more come to mind: Equivalence between two programs (especially in the context of meta- and generic programming), absence of heap allocations (for efficiency) or branching (for security).

While developers certainly have found ways to answer these questions before, as we see in Section 4, these solutions were ad-hoc and did not address the problem in full generality. By naming and describing this testing paradigm, we show a path to a more disciplined approach, and by following this path in the context of a specific programming language and implementation – Haskell and GHC – we show that inspection testing leads to concrete improvements to the quality of libraries, compilers and to increased developer productivity. The Haskell compiler GHC is a particularly good target to explore this idea, given its extensive and configurable set of non-local transformations, but inspection testing is applicable in all programming languages where the compiler transforms the code in non-trivial ways.

Our contributions are as follows.

- We identify that there exist relevant properties of programs that are commonly checked just by manual inspection and exhibit multiple cases of unnoticed regressions in released software (Sections 1 and 3.1.2).

- We propose *inspection testing* to automate this process. We give a precise definition for this paradigm in Section 2.

- Inspection testing is feasible: We describe the design choices of implementing inspection testing, and implement inspection testing for GHC (Sections 2.2 and 2.3).

- Inspection testing is effective: We describe how inspection testing is now used in practice to avoid regressions, to gain more confidence in the produced code and enable otherwise infeasible developments (Section 3).

- Inspection testing is beneficial: Users of inspection testing have a more solid idea of what they expect from their compiler, which has lead to improvements in the compiler (Section 3.1.4).

- Inspection testing scales: By automatically generating many inspection test obligations, we can achieve a test coverage that is infeasible with manual testing; we use this techinque to comprehensively check all the fusion-related promises in the text library (Section 3.3).

It might seem like inspection testing requires a genuine extension of the language and built-in support in the compiler, which would hamper adoption, but we can approximate the user experience of a real language extension using meta-programming and compiler plugins, as we explain in Appendix A.

## 2    Scope and design of inspection testing

Once the problem with manually inspecting compilation artifacts has been pointed out, the solution follows immediately: "Test it!" But conventional functional tests – whether unit testing, property-based or any other testing technique – cannot observe these properties. We need new testing paradigm, which we call *inspection testing*.

### 2.1    Definition

So what, precisely, is inspection testing – and what is it not? We propose the following definition:

> *Inspection testing is when a non-functional property of a compilation artifact of a specific piece of code is specified declaratively by the programmer and checked, during compilation, by the compiler.*

Let us unpack the crucial bits in this sentence, to see how inspection testing can be distinguished from other testing paradigms.

**non-functional** If the property was a functional one (e.g. "this list of integers is sorted") then checking this property would just be conventional testing.

**compilation artifact** This encompasses both intermediate representations internal to the compiler as well as the final compilation result. It excludes the actual source – checking purely syntactical properties is the realm of syntax checkers.

**specific piece of code** Properties that are expected to hold for all programs of a programming language need not be tested. They are either established by meta-theory (e.g. type safety) or ensured by the compilation toolchain (e.g. no unresolved symbols after linking).

**specified** Sometimes programmers read the intermediate programs and look for for "things that look strange". This is useful, but hard to automate – the compiler cannot implement the programmer's gut feeling. Therefore, inspection testing requires that the property is specified precisely.

**declaratively** If the compiler provides a plugin interface, or its source is available, then the programmer can just add new analyses to the compiler to check the intermediate code. We consider that, however, implementing a new static analysis. Inspection testing empowers programmers that do not want to, or cannot, extend their compiler.

***checked*** It may already be useful to be able to specify non-functional properties, e.g. in the documentation as a promise to the users of a library. But in order to speak of testing, the properties need to be checkable!

***by the compiler*** The checking should be performed as part of the normal compilation pipeline, rather than, say, using a separate tool. Otherwise, there is a risk that the result will not carry over to the actual compilation result.

## 2.2 Designing inspection testing

This abstract definition can guide us when we design inspection testing support in our favorite programming language and compiler. In this section, we show the steps that were necessary to bring inspection testing to Haskell. We expect that the same process can be followed for other programming languages.

1. The first step is to notice that we need inspection testing in the first place. Somewhat surprisingly, this is not obvious! As the examples in the introduction and in Section 3 show, developers care about such non-functional properties, but do not seem to expect to be able to test them.

2. Next, we have to find out what programmers are looking for when they inspect their compilation artifacts.

   Haskell programmers typically inspect their code at the stage of optimized Core. Core is GHC's intermediate language, an explicitly typed functional language, and most interesting transformations done by GHC apply to Core. With the flag -ddump-simpl, GHC prints the state of Core after all Core-to-Core optimizations have been applied.

   In the case of stream fusion, library authors would check the output of -ddump-simpl to see if "all values of type Text in a certain function have been eliminated." In other cases, programmers wonder whether "two functions, which are implemented differently, are actually the same to the compiler" (maybe in the variant "the same up to types") or they care about "this function does not allocate memory on the heap". Section 3 describes uses-cases where these questions come up.

   This step is naturally never complete, as developers will find new things they care about. Therefore, we need to keep the possibility of extending this list in mind.

3. We now take these verbal descriptions and cast them into a declarative, syntactical form for *test obligations*. In our case, we come up with four kind of obligations and gave them a syntax that is familiar to the Haskell programmer:

   - fun1 === fun2 expresses that after optimization, fun1 and fun2 have syntactically identical definitions (up-to renaming of variables).
   - fun1 ==- fun2 expresses that after optimization, fun1 and fun2 have definitions that differ only in types.
   - fun `hasNoType` ty expresses that fun's optimized definition does not mention the type ty.
   - noAllocations fun expresses that fun does not allocate memory on the heap.

4. This API allows the programmer to phrase their test obligations, which they now can declare within the source code. A straightforward way is to introduce a new top-level keyword (e.g. **inspect**) to the language:

   countChars :: ByteString −> Int
   countChars = length . toUpper . decodeUtf8

   **inspect** countChar `hasNoType` Text

   Again, different programming languages have different idioms for such extensions – a Python programmer might find it most natural use attributes to attach obligations to functions.

5. The compiler needs to be taught to recognize these declarations. So far we have assumed that we are free to extend the compiler and the syntax of the language, but that is not necessarily a realistic assumption. If it is not the case, or if we prioritize immediate adoption, we have to explore the existing option for extending the compiler.

   For Haskell, we found that with a combination of GHC plugins (which can hook into the optimization pipeline) and Template Haskell [Sheard and Jones 2002] we can provide a close approximation of the ideal of extending the syntax.

   In the end, the user writes some additional boiler-plate code to enable Template Haskell, to load the plugin, to import the module Test.Inspection that provides inspect and hasNoType, and – following the usual rules about Template Haskell – quotes the referenced function and type:

   {−# LANGUAGE TemplateHaskell #−}
   {−# OPTIONS_GHC -fplugin Test.Inspection.Plugin #−}

   **import** Test.Inspection

   countChars :: ByteString −> Int
   countChars = length . toUpper . decodeUtf8

   inspect ('countChars `hasNoType` ''Text)

   Appendix A describes in more detail how one can provide language extensions without modifying the compiler.

6. The compiler needs to actually perform the checks it was asked to do. Again, we face the question: Do we

```
module Test.Inspection where
inspect :: Obligation –> Q [Dec]          ––Test an obligation, and abort if it fails
inspectTest :: Obligation –> Q Exp        ––Test an obligation, and provide the result as a
data Result = Failure String | Success String   ––run-time value of type Result
type Obligation                           ––The type of test obligations
(===)              :: Name –> Name –> Obligation   ––Equivalence of top-level functions
(==-)              :: Name –> Name –> Obligation   ––Equivalence of top-level functions, up-to types
hasNoType          :: Name –> Name –> Obligation   ––Absence of a type in a function's body
hasNoAllocations :: Name –>             Obligation   ––Absence of allocations in a function's body
expectFailure :: Obligation –> Obligation     ––Declare that an obligation is expected to fail
```

**Figure 1.** The Test.Inspection module provided by inspection-inspection (slightly simplified)

have to actually modify the compiler, or are its extension mechanisms sufficient. For GHC, we can install our inspection testing check via the Core-to-Core plugin interface, which is meant to introduce additional optimization passes.

The actual checks are implemented as fairly straightforward analyses of the syntax tree. Obviously, this aspect of an inspection testing implementations will look very differently in other compilers or programming languages.

7. Finally, something needs to happen when the tests fail. There are three possibilities:

First, the compiler prints an error message explaining the failure and aborts the compilation. This is the default mode of our inspection testing support for Haskell: if the test

  inspect ('countChars `hasNoType` ''Text)

fails, the compiler prints the intermediate code of the function countChars and quits.

The compiler can simply dump the intermediate code onto the developer, who will have a similar experience than if he looked at the code manually. As a further refinement, the compiler can clearly highlight the interesting bits of the possibly very large code dump, based on the obligation that the user specified – in the example above, highlighting all values of type Text would enormously help the programmer in debugging this problem.

Second, the compiler prints a warning, but continues compilation. This mode is useful for example when, during a larger change to the code, the programmer wants to run the program and test its functionality before they address the regressions of non-functional properties. Our inspection-testing plugin for Haskell supports a keep-going flag to switch into this mode.

Just like with functional testing, it is useful to be able to mark certain test cases as "known to be failing." This can be used to document a known regression or to record intended future work. inspection-testing supports this.

A third variant is to continue compilation in any case, and make the result of the test available to the program being compiled. In the Haskell implementation, the user can write

```
test :: Result
test = $(inspectTest ('countChars `hasNoType` ''Text))
main = print test
```

to obtain a working program, which, when executed, prints the result of the inspection test. After the plugin checks the given obligation, it replaces the definition of test with a literal value of type Result, i.e.

```
test :: Result
test = Failure "Test failed: Function countChars …"
```

The advantage of this approach is that it allows inspection testing to be integrated seamlessly into an existing test suite of functional tests – see Section 3.1.3 for a real-world example.

### 2.3  Inspection testing in Haskell

The result of following these steps is the Haskell package inspection-testing, which is freely available on Hackage, the Haskell package repository. Its user-facing API is provided by the module Test.Inspection, summarized in Figure 1.

A user of the inspection-testing library imports the module Test.Inspection, ensures that the plugin in Test.Inspection.Plugin is loaded, enables the use of Template Haskell, and declares their test obligations using inspect. Let us elaborate the example from page 3 slightly and show the complete code:

```
{–# LANGUAGE TemplateHaskell #–}
{–# OPTIONS_GHC -fplugin Test.Inspection.Plugin #–}
module TextTests where
```

```haskell
import Test.Inspection

import Data.Text as T
import Data.Text.Encoding as E
import Data.ByteString (ByteString)

  -- A case of successful fusion:
toUpperBS :: ByteString -> String
toUpperBS = T.unpack . T.toUpper . E.decodeUtf8

  -- A failing case of fusion:
countChars :: ByteString -> Int
countChars = T.length . T.toUpper . E.decodeUtf8

inspect ('toUpperBS `hasNoType` ''T.Text)
inspect ('countChars `hasNoType` ''T.Text)
```

When we compile this file with GHC, the inspection plugin will test these two obligations and report the results. The second one fails, because its intermediate code still contains values of type Text. The plugin prints the Core of countChars, including the compiler-generated helper functions $wcountChars, as shown in Figure 2, and aborts compilation.

The Core representation is printed in the same ways as if the user had used -ddump-simpl, and the various flags modifying the output, such as -dsuppress-coercions, apply as expected. Our implementation of inspection testing does not yet highlight the interesting bits of the listing; until this is implemented, the user has to search for the occurrences of Text in the dumped data manually.

## 3 Inspection Testing in Action

Since the release of the inspection-testing library in November 2017, it has been adopted in a number of settings and put to good use. As evidence that inspection testing fills an existing but unrecognized need, and to inspire further adoption, we look at some of these uses.

### 3.1 The generic-lens promise

An incident that happened with the generic-lens library is a poster child for the need for inspection testing, and we will look at it in detail.

#### 3.1.1 What does generic-lens do?

The problem that generic-lens [Kiss 2017] is solving is the creation of *lenses*, using type-level computation and generic programming.

A lens can be thought of as a pointer to a field in a product data type. For example, given the data type definition

```haskell
data Employee = MkEmployee { name :: String, age :: Int }
```

we might be interested in the lens that points to the second field of Employee:

```haskell
ageLens :: Lens Employee Int
```

We can use this lens to read the field, set the field and modify the field. For example, if we have

```haskell
milton :: Employee
milton = MkEmployee { name = "Milton Waddams", age = 41 }
```

then we can use the operator (^.) to read Milton's age, i.e. milton ^. ageLens == 41, and use (^~) to set Milton's age, i.e. (ageLens ^~ 42) milton == MkEmployee "Milton Waddams" 42. Moreover, lenses are compositional, so we can chain them. And, since lenses are first class citizens in the language, we can store them in data structures and pass them to functions.

Lenses are very popular in the Haskell community, and many programmers will want a lens for every field in their data types. Of course, they can be implemented by hand:

```haskell
ageLens :: Lens Employee Int
ageLens f (MkEmployee name age) =
    fmap (\newAge -> MkEmployee name newAge) (f age)
```

Even without an explanation of the internals of the Lens type it is evident that writing such boiler plate code for every field in every data type quickly gets very tedious.

Therefore, most users use meta-programming to derive their lenses, using the function makeLenses provided by the lens package, and write something like

```haskell
data Employee = MkEmployee { _name :: String, _age :: Int }
makeLenses ''Employee
```

The makeLenses is a Template Haskell [Sheard and Jones 2002] function that will, at compile-time, produce the expected definitions of a lens for each of the two fields, named name and age. This solution works, but not everybody is happy with it: It requires to use underscore-prefixed names for the record fields, and the programmer has to decide upfront which types should be equipped with lenses. In fact, some developers shun the use of meta-programming completely.[2]

The ingenious generic-lens package, described by Kiss, Pickering, and Wu [2018], provides an alternative way to get lenses, based on generic programming and type-level computation. In this scheme, the user writes the data type as usual:

```haskell
data Employee = MkEmployee { name :: String, age :: Int }
    deriving Generic
```

Prompted by the **deriving** clause, the compiler makes the Employee type an instance of the Generic type class, which describes an isomorphism between the Employee type and a generic representation of it [Magalhães et al. 2010]. Without any further setup, the programmer can use the term field @"age" any time he needs a lens into the age field of the Employee type.

What happens conceptually under the hood when we use (field @"age" ~. 42) milton to update the age of milton?

---

[2] https://stackoverflow.com/a/10857227 lists a number of reasons

```
$ ghc -O TextTests.hs
[1 of 1] Compiling TextTests         ( TextTests.hs, TextTests.o )
TextTests.hs:19:1: toUpperBS `hasNoType` Data.Text.Internal.Text passed.
TextTests.hs:20:1: countChars `hasNoType` Data.Text.Internal.Text failed:
$wcountChars :: Addr# -> ForeignPtrContents -> Int# -> Int# -> Int#
$wcountChars =
                                      ... many lines of Core omitted ...
          case upperMapping @ Int (C# ww_sbsG) w_sbsD of _ { Done ->
          case unsafeFreezeByteArray# @ RealWorld ww_sbts w_sbt5 of _ {
            (# ipv5, ipv6 #) -> case ww_sbti of wild4
              { __DEFAULT -> (# ipv5, Text ipv6 0# wild4 #);
                                      ... many lines of Core omitted ...
countChars =
  \ (w_sbu1 :: ByteString) ->
    case w_sbu1 of _ [Occ=Dead] { PS ww_sbu4 ww_sbu5 ww_sbu6 ww_sbu7 ->
    case $wcountChars ww_sbu4 ww_sbu5 ww_sbu6 ww_sbu7 of ww_sbub
      { __DEFAULT -> I# ww_sbub }}

TextTests.hs: error:
    inspection testing unsuccessful
          expected successes: 1
          unexpected failures: 1
```

**Figure 2.** The inspection-testing library reports a a failed inspection test obligation

1. First, milton is converted to its generic representation, which for our purposes we can think of as the value ("*Milton Waddams*", 41) of type (String, Int) – the actual generic representation is a good deal more complicated.

2. Then, field uses type-level computation (i.e. type classes and type families [Chakravarty et al. 2005]) to find, in this generic data structure, the field with the requested name "age". Note that the argument to field is a *type-level argument* [Eisenberg et al. 2016], recognizable as such by the @ sign.

3. Once the field has been found, it is updated with the new value, and the generic value ("*Milton Waddams*", 42) is converted back to type Employee.

Of course, users of generic-lens do not have to know these details. They simply enjoy that they are able to address all fields as lenses, with almost no boilerplate code to write. Furthermore, they can use generic-lens to access fields by type or position and more.

### 3.1.2 What went wrong?

We have seen that generic-lens makes programming with lenses more convenient – but is the resulting code efficient? It seems that repeatedly converting data structures – which are likely much bigger than Employee – to their generic representation and back, cannot be as efficient as the hand-written or makeLenses-produced version, which work directly on the Employee type.

Nevertheless, in one paper [Kiss et al. 2017], the authors of generic-lens promise identical performance! They asked the Haskell compiler to show them the *optimized* intermediate code for both the hand-written version and the generically-derived version, and observed that – thanks to GHC's impressive optimization capabilities [Peyton Jones and Marlow 2002] – they were identical.

We were still skeptical, and when we tried to reproduce this, we could not: Even after optimization, the generically derived lens was converting between the concrete and the generic version.[3] It turned out that while the promise held for an earlier version, a small change to the internals of generic-lens got in the way of GHC's optimizations. Once identified, the authors could quickly fix the regression and release a new version of generic-lens.

### 3.1.3 Inspection testing to the rescue

In order to avoid such regressions in the future, the authors of generic-lens started to employ inspection testing: Their test suite[4] now contains a number of lenses over typical data types, defined both by hand and using the various generically derived variants. With the combinators provided by the inspection-testing library, they assert that the manual versions are, to the optimizing compiler, equivalent to the derived ones:

```
tests :: Test
tests = TestList $ map mkHUnitTest
    [ $(inspectTest $ 'fieldALensManual
                === 'fieldALensName)
```

---

[3]version 4.0.0.0, using GHC 8.2, reported at https://github.com/kcsongor/generic-lens/issues/13

[4]https://github.com/kcsongor/generic-lens/blob/master/test/Spec.hs

```
, $(inspectTest $ 'fieldALensManual
                === 'fieldALensType)
, $(inspectTest $ 'fieldALensManual
                === 'fieldALensPos)
, $(inspectTest $ 'subtypeLensManual
                === 'subtypeLensGeneric)
, $(inspectTest $ 'typeChangingManual
                === 'typeChangingGeneric)
, $(inspectTest $ 'typeChangingManual
                === 'typeChangingGenericPos)
, $(inspectTest $ 'typeChangingManualCompose
                === 'typeChangingGenericCompose)
```

… These assertions are integrated into their usual test suite harness based on HUnit [Herington and Hegel 2006], and are therefore checked by their continuous integration infrastructure upon every commit to the repository. This way, they can work on the internals of their library with confidence, as a change that would obstruct the crucial optimization would be detected immediately. On his blog, Csongor Kiss writes:

> *However, it happened multiple times during development that a small change (such as eta-reduction) broke the optimization.*
> *[The] excellent inspection-testing tool, which is now integrated into the automated test suite, is making sure that the optimization happens by automatically doing this comparison. This tool has been invaluable in ensuring the performance guarantees, without having to manually inspect the generated core after every single commit.*[5]

His coauthor Matthew Pickering writes in private communication:

> *[Before using inspection-testing], the first time that we upgraded the compiler version we noticed a regression which would have been prevented by having these tests. […] [inspection-testing] is very useful for ensuring the behavior across multiple GHC releases as well. Something which is quite hard to do manually.*

### 3.1.4 Improvements to the compiler

Matthew Pickering also reports that, thanks to Inspection Testing, he found several cases where the compiler failed to optimize the code to the expected extent and reported these bugs[6] to the GHC developers. Some have since been fixed have even led to improvements that are visible in GHC's benchmark suite "nofib" [Partain 1993].

The inspection-testing library helps users to think about the compiler's optimizations in a more structured way, to more clearly communicate their intent and to more easily identify the cause of regressions. We conjecture that this will lead them to hold the compiler to a higher standard and to demand optimizations that work reliably, rather than just on a best-effort basis.

## 3.2 Type-level programming

Oleg Grenrus's library vec defines sized-indexed lists, also known as vectors, with the following, unsurprising definition as an Generalized Algebraic Data Type [Cheney and Hinze 2003].

```
data Vec (n :: Nat) a where
    VNil :: Vec 'Z a
    (:::) :: a -> Vec n a -> Vec ('S n) a
```

Note that the quotation marks before Z and S promote these data constructors to the type level; this is not to be confused with the use of quotes in the context of Template Haskell, where they should be read as "name of."

The module Data.Vec.Lazy provides a collection of typical list-related functions (map, filter, zipWith etc.), defined naively using recursion. What sets vec apart from numerous other Haskell libraries providing a type for vectors is the module Data.Vec.Lazy.Inline, which provides the same API, but additionally guarantees that if the size of the vector is known at compile time, the recursion will be statically unrolled completely.

The test suite of vec uses inspection-testing to check that this is the case. For example, to test the zipWith functions, Grenrus defines two input vectors and the expected, statically computed output vector. He uses inspect both to check that the inlining variant really is evaluated statically, as well as to test that there indeed is a difference to the naive variant.

```
xs, ys, lhsNormal, lhsInline, rhsZipWith :: Vec N.Nat2 Int
xs          = 1 ::: 2 ::: VNil
ys          = 2 ::: 3 ::: VNil
lhsNormal   = L.zipWith (+) xs ys   -- Uses Data.Vec.Lazy
lhsInline   = I.zipWith (+) xs ys   -- Uses Data.Vec.Lazy.Inline
rhsZipWith  = 3 ::: 5 ::: VNil

inspect $                  'lhsInline   === 'rhsZipWith
inspect $ expectFailure $ 'lhsNormal =/= 'rhsZipWith
```

Under normal circumstances, GHC refrains from unrolling recursive functions like zipWith. Grenrus has to encode the recursion using type class instances to trick GHC in unrolling the recursion completely.[7] Such tricks are fiddly to get right, and employing them reliably is much more feasible with inspection testing. In fact, Grenrus writes:

> *I wouldn't claim that Data.Vec.Lazy.Inline stuff does really inline properly without confidence given by inspection-testing. Even writing that module in the first place would feel like a bad idea, given how much one would need to stare into the Core.*

---

[5] http://kcsongor.github.io/generic-lens/
[6] GHC tickets #14684, #14625 and #14688

[7] Grenrus names private communication with Andres Löh as the source for this trick, which is briefly described at
https://gist.github.com/phadej/1d04208a84f234778e309708f207e9af.

### 3.3 Comprehensive API checking

In the introduction we have seen that not all functions provided by the text library fuse as advertised. In particular, the example function pipeline consisting of the functions length, toUpper and decodeUtf8 failed to fuse. Naturally, we would like to know which of these three functions is the culprit? And what about the other 51 functions exported by Data.Text – do they fuse? Maybe only certain combination fuse? What about Data.Text.Lazy, which provides almost the same API, but with a different implementation?

It would be prohibitively labor-intensive to manually write down function pipelines for each of them, compile them and visually inspect the generated Core to answer all these questions. With the inspection-testing library, we can automate these steps: A metaprogram, using Template Haskell, composes each exported function with other, fusing functions and declares the corresponding inspection tests, using the inspectTest combinator. It also generates a thousand random pipelines, to check if certain combinations trigger unusual behavior. The generated program is compiled, the test obligations checked and the resulting programs compares the test results with the expected outcome, i.e. whether a pipelines where all functions are documented to fuse indeed fuses, and whether pipelines with a non-fusing function indeed does not fuse.

In module Data.Text, we investigated 63 functions. Of the 51 functions that are documented to fuse, 40 indeed fused, but 11 did not not (see Figure 3). Interestingly, of the 12 investigated functions that were not documented to fuse, two actually did – probably a documentation bug. The random testing revealed one interaction: The singleton functions makes some functions fuse that usually don't (e.g. center 42 . singleton will fuse with its consumer), but prevents other functions from fusing. We also inspected the module Data.Text.Lazy and found 6 functions that fail to fuse and 4 that fused unexpectedly.[8]

The 1000 random function pipelines are generated using combinators from the QuickCheck library Claessen and Hughes [2000]. Because of the phase separation – a metaprogram generates the test cases, then the compiler runs, then the resulting program processes the results – it is not straightforward to fully embed inspection tests into QuickCheck, for example to gain the benefits of shrinking counter-examples. But even in this rudimentary form we see that the benefits of random testing can apply in the realm of inspection testing.

### 3.4 Code synthesis

The Haskell package ghc-justdoit [Breitner 2018] relieves the programmer from implementing some functions where the functionality is already clear from the type.

---

[8]May 2018, using text-1.2.3.0 and GHC-8.4, test harness at https://github.com/nomeata/inspection-testing/tree/master/examples/text-api, reported at https://github.com/haskell/text/issues/202.

In Data.Text:
documented to fuse, but do not fuse:
unpackCString#, reverse, scanl1, scanr1, takeWhileEnd, dropWhileEnd, dropAround, strip, stripStart, stripEnd, length
not documented to fuse, but do fuse:
find, index

In Data.Text.Lazy:
documented to fuse, but do not fuse:
decodeUtf8, toCaseFold, scanl1, dropAround, strip, stripStart
not documented to fuse, but do fuse:
unfoldr, unfoldrN, find, index

**Figure 3.** Incorrectly documented fusion behaviour in the text library

For example, the programmer can write

```
myBind :: (r –> Either e a) –> (a –> (r –> Either e b))
                            –> (r –> Either e b)
myBind = justDoIt
```

and the compiler finds the intended implementation.

The function justDoIt itself is the only method of a type class defined a

**class** JustDoIt a **where** justDoIt :: a without any instances. The "magic" works due to a compiler plugin that comes with the ghc-justdoit package and hooks into the GHC type checker. It finds unsolved constraints of the form JustDoIt t, interprets the type t as a formula in intuitionistic propositional logic and tries to prove them using LJT proof search [Dyckhoff 1992] – much like the well-known command-line tool djinn by Augustsson [2008]. If it finds such a proof, it interprets it as a Core expression of type t and passes it to the compiler as the "instance" to satisfy this constraint.

The development of ghc-justdoit was significantly more pleasant thanks to inspection-testing: it allowed us to simply state the expected code and have that checked, instead of tediously having to ask GHC to dump the Core and manually inspecting that:

```
myBindManual :: (r –> Either e a) –> (a –> (r –> Either e b))
                                  –> (r –> Either e b)
myBindManual m1 m2 r =
    case m1 r of Left e   –> Left e
                 Right a –> m2 a r
inspect $ 'myBind === 'myBindManual
```

### 3.5 Erasure in proof-carrying code

Liquid Haskell, which brings refinement types to Haskell [Vazou et al. 2014], can assist the programmer in deriving efficient code from high-level specification, by a series of small, verified equational reasoning steps [Vazou et al. 2018]. One

```
{-@ rev' :: xs:[a] -> ys:[a] -> {zs:[a] | zs == reverse xs ++ ys} @-}
rev' :: [a] -> [a] -> [a]
rev' []     ys =    reverse []    ++ ys
                ==. [] ++ ys
                ==. ys
rev' (x:xs) ys =    reverse (x:xs) ++ ys
                ==. (reverse xs ++ [x]) ++ ys
                ==. reverse xs ++ ([x] ++ ys)
                ?   assocP (reverse xs) [x] ys
                ==. rev' xs ([x] ++ ys)
                ==. rev' xs (x:([] ++ ys))
                ==. rev' xs (x:ys)
{-@ reverseOpt :: xs:[a] -> {v:[a] | v == reverse xs } @-}
reverseOpt :: [a] -> [a]
reverseOpt xs =     reverse xs
                ==. reverse xs ++ []
                ?   leftIdP (reverse xs)
                ==. rev' xs []
```

**Figure 4.** Verified program derivation using Liquid Haskell

example in that paper is the derivation of an efficient list reversal, where every step of the derivation is verified against the naive implementation of reverse, as seen in Figure 4.

At first glance, the definitions of rev' and reverseOpt in the figure form a *proof* of the correctness of the efficient implementation, which we would then write as

```
rev' :: [a] -> [a] -> [a]
rev' []     ys = ys
rev' (x:xs) ys = rev' xs (x:ys)

reverseOpt :: [a] -> [a]
reverseOpt xs = rev' xs []
```

But in fact, the functions in the figure can directly be used as the *implementation*, because the proof combinator (==.) is cleverly set up to return its second argument.

Given that performance was the motivation of this derivation, using the proof as the implementation is only attractive if it does not come with a performance penalty. One might expect that GHC detects that only the expression in the last line of each derivations matters and optimize the other expressions away. Indeed, the authors promise that the code with the proof steps and the code without the proof steps are identical after optimizations, and they do that with confidence thanks to inspection-testing.

## 4 Related Work

The need for testing of non-functional properties, even if not recognized as such, has certainly itched developers in the past, and some scratched this itch in a way that we can consider to be examples of inspection testing.

### 4.1 Compiler test suites

Compilers come with test suites of impressive size. GCC ships with over 42,000 test files, LLVM runs about 40,000 tests and the Haskell compiler GHC has accumulated the still sizable number of over 6,000 test cases. The test cases typically come with a small program and simply check whether it compiles – or fails to compile with the expected error message. Other test cases run the compiled program to check if it behaves as expected. But all three compiler's test suites sport a few test cases that explicitly investigate the intermediate or assembly code, by checking its textual representation for the presence of absence of certain strings or regular expressions.

One can, if so inclined, debate whether this is inspection testing: After all, when developing a compiler, the compiler is "just" the product, and its optimizations are its functional features, so these tests can be considered *functional* properties. In any case, a compiler that offers inspection testing support to their users will also have better tools to use in their own test suite.

### 4.2 FileCheck

The LLVM test suite relies heavily on the FileCheck program,[9] also provided by LLVM. It can be thought of as a more powerful variant of grep: it reads search strings and patterns from a file – commonly the program source file – and makes sure they occur, in order, in the text passed via its standard input. To test the code from the Introduction this way we first add comments with FileCheck directives to the source file:

```
module FileCheckTest where
import Data.Text as T
import Data.Text.Encoding as E
import Data.ByteString (ByteString)

-- CHECK: countChars
-- CHECK-NOT: Text
countChars :: ByteString -> Int
countChars = T.length . T.toUpper . E.decodeUtf8
```

We pass this file both to GHC for compilation, and to FileCheck as the specification, and observe that it reports the error, as expected:

```
$ ghc -O -ddump-simpl FileCheckTest.hs |
    FileCheck FileCheckTest.hs
<stdin>:32:4: error: CHECK-NOT: string occurred!
 @ Text
 ^
FileCheckTest.hs:7:15: note: CHECK-NOT:
    pattern specified here
-- CHECK-NOT: Text
```

The generic, text-based approach taken by FileCheck is very flexible and easy to get started with, and other projects – such as the Rust compiler rustc and the D compiler LDC – have adopted it in their test suites.

---

[9] https://llvm.org/docs/CommandGuide/FileCheck.html

This example shows that it is possible to implement inspection testing using FileCheck. We think, however, that a language-specific implementation that works on the actual AST of the intermediate language provides a better experience. In particular, text matching is fragile and imprecise: the above test would fail also if a string literal, module name or function name contained the string "*Text*", whereas the declarative inspection testing obligation countChars `hasNoType` ''Text has a precise semantic meaning. Some more complicated properties – absence of allocation, alpha-equality of code fragments – might even be impossible to check using a purely textual tool.

### 4.3 Making non-functional properties functional

In our definition of inspection testing, we emphasize the distinction between functional properties, which can be tested conventionally, and non-functional properties, which require a separate paradigm. However, the border between these two can be a bit blurry, and may even be shifted to make non-functional properties conventionally testable, as the following two examples – taken from Haskell – show:

***Testing laziness***  The evaluation order in a pure programming language is, at first glance, a non-functional property: It should not affect the semantics of the program if the compiler decides to reorder evaluations. A developer might still want to know whether each iteration allocates a new suspended computation, i.e. a *thunk*, or if – thanks to a clever compiler optimization – the argument is evaluated in each iteration and passed as a value. A check "this function does not allocate thunks" is clearly within the realm of inspection testing.

But in not-fully-pure languages with side effects – whether its unrestricted IO like in ML, or merely exceptions like in Haskell – we can observe evaluation order to some extent, and suddenly it becomes a functional property. Indeed, the test suite of the Haskell library containers [library maintainers 2007], which provides map data structures in variants that differ in whether their operations evaluate the stored values eagerly or lazily, tests these properties as part of its property tests: using QuickCheck [Claessen and Hughes 2000], it crafts inputs that raise exceptions upon evaluation, and checks if executing the map operations triggers these exceptions.

The recently published library StrictCheck by Foner, Zhang, and Lampropoulos [2018] facilitates such testing of laziness and strictness properties: it allows the programmer to give precise specifications of the expected evaluation behavior of a function, and tests these specifications, at run-time, using randomized testing. It uses unsafe features of the Haskell runtime to observe the evaluation behavior in ways that are not possible in "normal" Haskell code.

***Making fusion visible***  List fusion [Peyton Jones et al. 2001] combines a function that produces a list with the function that consumes the list into a single recursive function, where no intermediate lists are produced. This optimization is a poster-child for something that we need inspection testing to check reliably, because it has no effect on the semantics of the program.

We can, however, abuse the rewrite rule machinery that implements list fusion to make it observable. Recall the central rule of list fusion,

```
{-# RULES "fold/build"
    forall k z (g::forall b. (a->b->b) -> b -> b).
    foldr k z (build g) = g k z #-}
```

which combines a list producer (expressed in terms of the combinator build) with a list consumer (expressed in terms of the combinator foldr). Together with some additional rules this achieves that the expression map f (map g xs) fuses to map (f . g) xs, because map can be expressed in terms of foldr and build. In contrast, sort (map g xs) does not fuse, because sort is not expressed as a simple right-fold over its argument.

The list-fusion-probe package [Breitner 2014] allows the programmer to make this difference observable, by marking spots where they expect fusion to happen with the special function fuseThis. This function is defined to simply crash:

```
fuseThis :: [a] -> [a]
fuseThis = error "fuseThis: List did not fuse"
```

It comes with a rewrite rule, which is a copy of the "*fold/build*" rule, but allows the fuseThis function to appear between the list consumer and the list producer, and removes it from the program:

```
{-# RULES "fold/fuseThis/build"
    forall k z (g::forall b. (a->b->b) -> b -> b) .
    foldr k z (fuseThis (build g)) = g k z #-}
```

The effect is that the expression map f (fuseThis (map g xs)) will fuse to map (f . g) xs and run just fine. In contrast, sort (fuseThis (map g xs)) will not fuse, the call to fuseThis remains and alerts the programmer, at runtime, that some expected fusion did not happen.

We see that some properties that – in the mindset of the developer – are not functional properties, can still be tested conventionally if one can make – maybe using "unsafe" tricks – the program behave observably differently. This provides an alternative when no inspection testing implementation is available, or if it is not able to express the desired properties.

### 4.4 Testing is good, control is better

In a way, any kind of testing is just a work-around for insufficiently expressive languages: You only need to test for proper memory allocation usage, e.g. using valgrind [Nethercote and Seward 2007], if your programming language is not memory safe. You only need to test if your functions return the right data structures if your programming language is not type safe. You only need to write property tests if your language does not allow you to verify your programs.

Inspection testing is no different: If our language and compiler allows us to instruct the compiler to do something, then we do not have to test this. Many use cases for inspection testing in this paper are related to generic programming, and ensuring that certain transformations happen at compile-time rather than at run-time.

With a meta-programming system with a clear separation of phases, such as Scala LMS [Rompf and Odersky 2012], the code is explicit in when which computation happens, and we do not need to test that it happened by the time the meta-program has finished. Note that we still might want to use inspection testing to investigate what the compiler then does to code created by metaprograms.

Similarly, the Haskell compiler GHC provides a number of pragmas and magic functions that give clear instructions to the optimizer: for example, if a function is annotated with an INLINE pragma, then we should be able to rely on the compiler indeed inlining the function, at least within the constraints documented in the user manual.

## 5  Future work

The problems that inspection testing solves are present in any language with highly optimizing compilers, meta-programming or type-level programming. An example from Scala is curryhoward by Winitzki [2001], a code inference plugin similar to the Haskell package ghc-justdoit discussed in Section 3.4: The term def f1[X] = ofType[X => X] infers code for f1 with the given type. Its test suite wants to test if f1 *really is* the identify function, but can only test whether f1 *behaves* like the identity.

In Section 3.3 we have seen that property based testing and random testing apply equally well to inspection-testing. To fully exploit this potential, we need to find a way to feed the results of the tests back to the generator, for example, for shrinking. This can be achieved if the produced program writes the test results to a file that the metaprogram can read, and repeating the compilation as often as necessary.

The set of obligations that our inspection testing implementation can handle is not set in stone, and needs to be improved based on user demand. Additionally, the syntax to describe obligations can be expanded: we envision a DSL to describe properties, which can provide combinators to address specific code fragments ("the inner loop of function foo").

Currently, inspection-testing is provided as a stand-alone plugin, which is great during this early stage of development and adoption, but has the cost of slightly less convenient syntax. Once it has matured some more, we can address the question of whether this should become a built-in feature of the compiler.

## 6  Conclusion

There is a real and unrecognized need for inspection testing out there, as shown by multiple incidents where documented promises about released software did not hold. Once this problem has been pointed out, it is clear what the solution is. Implemented for Haskell, it has seen swift adoption by the target audience and already improved the quality of our software libraries and compilers. We are confident that both the problem and the solution applies to other programming languages, and that inspection testing will have a positive impact there as well.

## Acknowledgments

We would like to thank Benjamin Pierce and Leo Lampropoulos for their helpful comments. This material is based upon work supported by the National Science Foundation under Grant No. 1319880 and Grant No. 1521539.

## References

Lennart Augustsson. 2008. djinn: Generate Haskell code from a type. http://hackage.haskell.org/package/djinn. (2008).

Joachim Breitner. 2014. list-fusion-probe: testing list fusion for success. http://hackage.haskell.org/package/list-fusion-probe. (2014).

Joachim Breitner. 2015. Call Arity. In *Trends in Functional Programming (TFP) 2014 (LNCS)*, Vol. 8843. Springer. https://doi.org/10.1007/978-3-319-14675-1_3

Joachim Breitner. 2017. The sufficiently smart compiler is a theorem prover (extended abstract). In *IFL*.

Joachim Breitner. 2018. ghc-justdoit: A magic typeclass that just does it. http://hackage.haskell.org/package/ghc-justdoit. (2018).

Manuel M. T. Chakravarty, Gabriele Keller, Simon L. Peyton Jones, and Simon Marlow. 2005. Associated types with class. In *POPL*. ACM, 1–13. https://doi.org/10.1145/1040305.1040306

James Cheney and Ralf Hinze. 2003. *First-Class Phantom Types*. Technical Report. Cornell University.

Koen Claessen and John Hughes. 2000. QuickCheck: a lightweight tool for random testing of Haskell programs. In *ICFP*. ACM, 268–279.

Iavor S. Diatchki. 2015. Improving Haskell types with SMT. In *Haskell Symposium*. ACM, 1–10. https://doi.org/10.1145/2804302.2804307

Roy Dyckhoff. 1992. Contraction-Free Sequent Calculi for Intuitionistic Logic. *Journal of Symbolic Logic* 57, 3 (1992), 795–807. https://doi.org/10.2307/2275431

Richard A. Eisenberg, Stephanie Weirich, and Hamidhasan G. Ahmed. 2016. Visible Type Application. In *ESOP (LNCS)*, Vol. 9632. Springer, 229–254. https://doi.org/10.1007/978-3-662-49498-1_10

Conal Elliott. 2017. Compiling to categories. *PACMPL* 1, ICFP (2017), 27:1–27:27. https://doi.org/10.1145/3110271

Kenneth Foner, Hengchu Zhang, and Leonidas Lampropoulos. 2018. Keep your Laziness in Check. (2018).

Andy Gill and Simon Marlow. 2001. Happy: The Parser Generator for Haskell. (2001). http://haskell.org/happy/

Dean Herington and Simon Hegel. 2006. HUnit: A unit testing framework for Haskell. http://hackage.haskell.org/package/HUnit. (2006).

Csongor Kiss. 2017. generic-lens: Generic data-structure operations exposed as lenses. http://hackage.haskell.org/package/generic-lens. (2017).

Csongor Kiss, Matthew Pickering, and Toby Shaw. 2017. Deriving lenses using generics. In *IFL*.

Csongor Kiss, Matthew Pickering, and Nicolas Wu. 2018. Generic Deriving of Generic Traversals. In *ICFP*. ACM.

Chris Lattner and Vikram S. Adve. 2004. LLVM: A Compilation Framework for Lifelong Program Analysis & Transformation. In *CGO*. IEEE Computer Society, 75–88.

The Haskell library maintainers. 2007. containers: Assorted concrete container types. http://hackage.haskell.org/package/containers. (2007).

José Pedro Magalhães, Atze Dijkstra, Johan Jeuring, and Andres Löh. 2010. A generic deriving mechanism for Haskell. In *Haskell Symposium*. ACM, 37–48. https://doi.org/10.1145/1863523.1863529

Nicholas Nethercote and Julian Seward. 2007. Valgrind: A Framework for Heavyweight Dynamic Binary Instrumentation. In *Programming Language Design and Implementation (PLDI)*. ACM. https://doi.org/10.1145/1273442.1250746

Bryan O'Sullivan and Tom Harper. 2017. text: An efficient packed Unicode text type. http://hackage.haskell.org/package/text-1.2.2.2. (2017).

Will Partain. 1993. The nofib Benchmark Suite of Haskell Programs. In *Functional Programming 1992 (Workshops in Computing)*. Springer. https://doi.org/10.1007/978-1-4471-3215-8_17

Simon Peyton Jones. 1992. Implementing Lazy Functional Languages on Stock Hardware: The Spineless Tagless G-Machine. *Journal of Functional Programming* 2, 2 (1992). https://doi.org/10.1017/S0956796800000319

Simon Peyton Jones and Simon Marlow. 2002. Secrets of the Glasgow Haskell Compiler Inliner. *Journal of Functional Programming* 12, 5 (2002). https://doi.org/10.1017/S0956796802004331

Simon Peyton Jones, Andrew Tolmach, and Tony Hoare. 2001. Playing by the rules: rewriting as a practical optimisation technique in GHC. In *Haskell Workshop*.

Tiark Rompf and Martin Odersky. 2012. Lightweight modular staging: a pragmatic approach to runtime code generation and compiled DSLs. *Commun. ACM* 55, 6 (2012), 121–130. https://doi.org/10.1145/2184319.2184345

Ilya Sergey, Dimitrios Vytiniotis, Simon L. Peyton Jones, and Joachim Breitner. 2017. Modular, higher order cardinality analysis in theory and practice. *Journal of Functional Programming* 27 (2017), e11. https://doi.org/10.1017/S0956796817000016

Tim Sheard and Simon L. Peyton Jones. 2002. Template meta-programming for Haskell. *SIGPLAN Notices* 37, 12 (2002), 60–75. https://doi.org/10.1145/636517.636528

Martin Sulzmann, Manuel M. T. Chakravarty, Simon L. Peyton Jones, and Kevin Donnelly. 2007. System F with type equality coercions. In *TLDI*. ACM, 53–66. https://doi.org/10.1145/1190315.1190324

Niki Vazou, Joachim Breitner, Will Kunkel, David Van Horn, and Graham Hutton. 2018. Theorem proving for all. (2018). submitted to Haskell'18.

Niki Vazou, Eric L. Seidel, Ranjit Jhala, Dimitrios Vytiniotis, and Simon L. Peyton Jones. 2014. Refinement types for Haskell. In *ICFP*. ACM, 269–282. https://doi.org/10.1145/2628136.2628161

Sergei Winitzki. 2001. curryhoward: Automatic code generation for Scala functions and expressions via the Curry-Howard isomorphism. https://github.com/Chymyst/curryhoward. (2001).

## A  Faking a language extension

In Section 2.2 we describe the steps to take to bring inspection testing to an existing programming language. It requires adding new syntactic constructs and reacting to them during the compilation process. It seems that implementing this would require changing the compiler itself. Most programming languages have implementations that are Free Software, so it is possible to fork them and create a custom version with support for inspection testing. But it is questionable whether possible users out there would go through the hassle of using a different compiler, so this approach would greatly hamper adoption.

Luckily, in many cases, we will be able to approximate the ideal of a seamless language extension, namely when the language supports meta-programming (to give the illusion of new syntax) and some form of compiler plugins.

We demonstrate this idea with the Haskell compiler GHC and describe – in this mostly self-contained section – how we can use (or abuse) some features of GHC to emulate a language extension.

The premise is that we aim to add functionality to the Core phase of the compiler – this could be a check like inspection-testing, or some active modification, such as optimizing DSLs like Elliott's "Compiling to Categories" [2017]. Additionally, we assume that this functionality needs some form of user input that we want to include in the source level – if no input is necessary, e.g. for a plain new optimization pass, then GHC's plugin interface is perfectly adequate. Three problems need to be solved for this:

1. Which syntax does the user use?
2. How is the user input transported from the source code to the Core phase?
3. How is the user input kept alive through existing optimizations, in particular dead code elimination?

We start with the second question, as its answer affects how we solve the other two, after we have recapitulated some relevant bits of GHC's architecture.

### A.1  GHC's compilation pipeline

The Haskell compiler GHC processes a module in the following steps:

1. The file is loaded from disk and parsed. The *parser*, which is implemented using the parser generator happy [Gill and Marlow 2001], is highly configurable, because GHC supports a large number of opt-in language extensions that extend the syntax, but it is not extensible: All these language extensions are hard-coded in the compiler.

   The result of this parsing pass is an abstract syntax tree of the full source Haskell language which faithfully reflects the syntactic sugar used by the programmer (type HsExpr and friends).

2. The file is processed by the renamer and the type-checker. Conceptually these are two different phases, but they actually work in lock-step.

   Variable names in the output of the parser are essentially strings (type RdrName). The *renamer* resolves these strings to actual names, which unambiguously point to their binding site (type Name). Imported names are qualified with the package identifier and module name, and all names receive a unique number, which from now on is used to decide if two names refer to the same thing. The textual name is preserved, but only used to make debugging output more readable.

   The *type-checker* traverses the syntax tree, infers the type of expressions where necessary, and checks that the program is well-typed. It also resolves type class constraints. It enriches the source code AST with the result of type inference, essentially elaborating it to a typing judgment deriving tree (using type HsWrapper). It also replaces names (type Name) with identifiers (type Id and Var) that carry additional information; in particular, the type of the variable.

   GHC allows plugins to hook into the type checker and act on unresolved constraints. We will discuss this alternative way of extending GHC briefly in Section 3.4.

3. Once the program has been successfully type-checked, the *desugarer* transforms the very complex and rich source AST (type HsExpr) into GHC's much smaller intermediate language Core (type CoreExpr). Core is an explicitly typed functional programming language based on System $F_C$ [Sulzmann et al. 2007]. Many features of Haskell are dissolved at this point: All the syntactic sugar, nested pattern matches, type classes and GADTs are all expressed in terms of Core's more primitive features.

4. GHC runs a number of optimization passes on the core AST. This includes the general-purpose simplifier [Peyton Jones and Marlow 2002], as well as more specialized analysis like the demand and strictness analyzer [Sergey et al. 2017] or Call Arity [Breitner 2015].

   A GHC plugin can modify the list of optimization passes to run; in particular, it can install new passes at any point in the list. Such a pass is simply a monadic function of type ModGuts –> CoreM ModGuts, where ModGuts contains all the information about the current module (in particular, the Core AST), and CoreM is a monad that gives the plugin access to some GHC internals, e.g. to allow looking up names in the environment, but also full access to the IO monad.

5. The stages following Core are less relevant for our purpose: The optimized Core is transformed to the untyped functional language STG [Peyton Jones 1992], mildly optimized there, then compiled to the portable assembly language C−−, with more optimizations applied here. Eventually, GHC either creates native assembly code directly, or it creates LLVM's intermediate representation and uses LLVM [Lattner and Adve 2004] to finish compilation.

### A.2 Transporting additional information in the AST

Given this compiler architecture, we need to find a way to include the user's input in the AST in a way that survives the renamer, typechecker, desugarer and Core-to-Core optimizations. Several options come to mind:

#### A.2.1 Annotations

One obvious way is the "official" way, using GHC's support for *source annotations*. Any value of a type with an instance of the type class Data can be used as a source annotation, and can be attached to a top-level definition, type declaration or the module itself.

The expression itself is evaluated at compile time, much like a Template Haskell expression. In particular, we can include a reference to other names in the module with a quoted name.

For the inspection test mentioned in the introduction, where we want to check the absence of a certain type in a definition, we would define a data type like

```
import Language.Haskell.TH.Syntax (Name)
data HasNoTypeAnn = HasNoType Name deriving Data
```

to be used as an annotation of countChar:

```
countChars :: ByteString –> Int
countChars = length . toUpper . decodeUtf8
{–# ANN countChars (HasNoType ''Text) #–}
```

The compiler serializes the annotation to a binary representation (using a generic scheme based on the Data instance) tags it with its type (using the Typeable instance, which GHC creates for all types) and stores it in the mg_anns field of the type ModGuts. Our plugin can find it there, deserialize it to get a the value of type HasNoTypeAnn, and act on it. In our example, the argument to HasNoType is still the Template Haskell name of Text, which is not very useful at the Core stage, but we can use GHC's function thNameToGhcName to convert this into GHC's internal Name.

#### A.2.2 Magic functions

Annotations work great when it suffices to mark top-level declarations, but unfortunately, GHC does not support annotating local declarations or whole subexpressions. If that is needed, then "magic functions", albeit a bit hackish, can go a long way.

The idea is to define a function that purely serves as a marker in the source code, but has no (interesting) functionality of its own. As an example, let us allow the user to mark subexpressions where they expect no allocations. We would define a function noAllocations with a flexible type:

```
module Test.NoAllocations where
noAllocations :: a –> a
noAllocations x = x
{–# NOINLINE noAllocations #–}
```

Now our plugin can traverse the Core AST and search for occurrences of noAllocations ty e (at the level of Core, noAllocations has two parameters: The type a, and the value of type a). It checks the subexpression e for the desired property, and finally replaces the whole expression by e, so that in the end, no calls to noAllocations remain. We have defined noAllocations so that even if the plugin is not run and the calls to noAllocations remain, the semantics of the program is unaffected.

It is crucial to mark this function with a NOINLINE pragma, as otherwise, the simplifier would already replace noAllocations ty e with e before our plugin has a change to look at it.

When following this path one has to remember that GHC will still happily move code in and out of noAllocations, e.g. it might transform the expression noAllocations (fib (42 + 32)) to the expression **let** x = 42 + 23 **in** noAllocations (fib x).

The GHC plugin ghc-proofs [Breitner 2017], a precursor to inspection-testing which uses GHC's optimizer as a simple but convenient theorem prover, is using this method.

#### A.2.3 Rewrite rules

A third alternative is to (ab)use Haskell's rewrite rules [Peyton Jones et al. 2001]. Rewrite rules are specified in Haskell source syntax, then renamed, type-checked and desugared as usual and finally used by the Core simplifier to rewrite the code. For example, the rule

```
{–# RULES "map/map"
    forall f g xs. map f (map g xs) = map (f . g) xs #–}
```

fuses a sequence of calls to map to a single one. When we want to use rewrite rules to transmit information, we obviously don't want the simplifier to actually apply it. We achieve that by including an otherwise unused, special function on the left-hand side of the rule.

As an example, let us give the user the ability to ask for two expressions to be compared. We would define

```
isSameCode :: a –> a
isSameCode x = x
{–# NOINLINE isSameCode #–}
```

so that the user can write

```
test1, test2 :: [Bool] –> [Bool]
test1 xs = map not (map not xs)
test2 xs = xs

{–# RULES "inspect" isSameCode test1 = test2 #–}
```

Our plugin finds all rules that mention isSameCode and compares the argument of isSameCode on the left-hand side with the expression on the right-hand side.

The advantage of this approach over annotations is that it allows the user to write Haskell source expressions, while the plugin receives the corresponding Core – there is no need to worry about converting Template Haskell names to Core names, for example. On the other hand, it is an abuse of a feature that is intended for something else.

### A.3 Keeping the information alive

In all three variants we have to worry about keeping the user's information alive until it reaches our plugin and also about removing it afterwards. GHC's optimizer removes code that it thinks is not used. It considers the following functions as used:
- functions exported from the current module or, if there is no export list, all top-level functions,
- functions used in type class instances,
- functions mentioned on either side of a rewrite rule,
- functions whose quoted name is mentioned in any Template Haskell splice or annotation
- and of course all functions used by such functions.

Note that annotating a function on its own does not keep it alive! For example, consider the following module, which exports nothing:

```
module Test () where
test1, test2 :: [Bool] –> [Bool]
test1 xs = map not (map not xs)
test2 xs = xs
{–# ANN test1 (const "daedal") 'test2 #–}
```

The compiler will quickly drop test1 (together with its annotation!), but it will keep test2 around. It suffices that test2 is *mentioned* in the annotation expression, even though the compiler evaluates this expression to just "*daedal*" before storing it.

This has consequences for the design of our plugin: In any scheme that uses quoted mentions in Template Haskell splices or annotation, the mentioned functions stay alive until our plugin runs (good!), but also continue to stay alive afterwards (possibly bad).

Rewrite rules offer more flexibility here: They keep referenced code reliably alive, but the plugin may choose to remove the rewrite rules, which then allows GHC to remove the referenced code.

### A.4 Presenting a nice user interface

The most suitable option to include the user input in the AST, and to keep it alive, might not be the most friendly way for the user. A "magic" function is probably sufficiently intuitive; encoding the information in carefully crafted rewrite rules is certainly less so.

This problem can be overcome by careful use of Template Haskell, where a Template Haskell splice produces the necessary definitions, annotations and rewrite rules to convey the user's intent. For example, instead of asking the user to write the HasNoType annotation as shown in in Appendix A.2.1, we provide a function

```
assertHasNoType :: Name –> Name –> Q [Dec]
assertHasNoType fun ty = do
    ann <– liftData (HasNoType ty)
    return [ PragmaD (AnnP (ValueAnnotation fun) ann) ]
```

To the user, using this function feels almost as if it were a new-top-level declaration form:

```
assertHasNoType 'countChars ''Text
```

The effect of this function is simply to generate the annotation seen in Appendix A.2.1. A side-effect of this is that countChars is kept alive, as explained in the previous section.

This is the way we chose for inspection-testing: The user uses the inspect function to declare his test obligations, which in turn generates the actual annotations to transport the obligations to the plugin.

### A.5 Alternative approaches

We would like to point out other ways to extend GHC.

***Type-checker plugins*** A mostly orthogonal approach to extend GHC is to use *type-checker plugin*. As their name indicates, these plugins hook into the type-checker phase, and get the chance to solve constraints that the compiler did not solve on its own. If applicable, this approach can provide a very seamless integration of new features into the language.

This has been used, for example, to discharge type-level calculations to an SMT solver [Diatchki 2015], or to add type-driven code synthesis to GHC (see Section 3.4).

***New front-end programs*** It may be that even with all the tricks we presented, it is not possible to implement our extension within GHC, and we have to implement our own "compiler." Even in that case, we can re-use large parts of GHC, because all of its functionality is available as the Haskell library ghc, often referred to as "GHC-the-library."

This approach was taken by Liquid Haskell [Vazou et al. 2014], where the user needs to be able to add refinement type signatures not only to top-level declarations, but also to local functions. Furthermore, Liquid Haskell needs to preserve more information about the source code, such as code line numbers, until after the desugarer has converted the Haskell program into GHC Core. The result is the syntax shown in Section 3.5: The file can still be compiled by GHC as usual, but when processed by the liquid command, the special {–@ … @–} pragmas are interpreted and used in the refinement type checking pass, which works on GHC core.

In the long run it would be desirable if GHC would offer the necessary hooks so that a feature like refinement types can be implemented completely as a plugin.